\pdfoutput=1
\RequirePackage{ifpdf}
\ifpdf 
\documentclass[pdftex]{sigma}
\else
\documentclass{sigma}
\fi

\begin{document}

\allowdisplaybreaks

\renewcommand{\thefootnote}{$\star$}

\renewcommand{\PaperNumber}{037}

\FirstPageHeading

\ShortArticleName{Quantum Integrals for a~Semi-inf\/inite~$q$-Boson System}

\ArticleName{Quantum Integrals for a~Semi-Inf\/inite $\boldsymbol{q}$-Boson System\\
with Boundary Interactions\footnote{This paper is a~contribution to the Special Issue on Exact Solvability and Symmetry
Avatars in honour of Luc Vinet.
The full collection is available at
\href{http://www.emis.de/journals/SIGMA/ESSA2014.html}{http://www.emis.de/journals/SIGMA/ESSA2014.html}}}

\Author{Jan Felipe VAN DIEJEN~$^\dag$ and Erdal EMSIZ~$^\ddag$}

\AuthorNameForHeading{J.F.~van Diejen and E.~Emsiz}

\Address{$^\dag$~Instituto de Matem\'atica y F\'{\i}sica, Universidad de Talca, Casilla 747, Talca, Chile}
\EmailD{\href{mailto:diejen@inst-mat.utalca.cl}{diejen@inst-mat.utalca.cl}}

\Address{$^\ddag$~Facultad de Matem\'aticas, Pontificia Universidad Cat\'olica de Chile,\\
\hphantom{$^\ddag$}~Casilla 306, Correo 22, Santiago, Chile}
\EmailD{\href{mailto:eemsiz@mat.puc.cl}{eemsiz@mat.puc.cl}}

\ArticleDates{Received February 04, 2015, in f\/inal form April 30, 2015; Published online May 06, 2015}

\Abstract{We provide explicit formulas for the quantum integrals of a~semi-inf\/inite~$q$-boson system with boundary
interactions.
These operators and their commutativity are deduced from the Pieri formulas for a~$q\to 0$ Hall--Littlewood type
degeneration of the Macdonald--Koornwinder polynomials.}


\Classification{33D52; 81R50; 81T25; 82B23}

\renewcommand{\thefootnote}{\arabic{footnote}} \setcounter{footnote}{0}

\section{Introduction}

The~$q$-boson system~\cite{bog-ize-kit:correlation,sas-wad:exact} constitutes an integrable~$q$-deformed lattice
regularization of the quantum nonlinear Schr\"odinger
equation~\cite{dor:orthogonality,kor-bog-ize:quantum,lie-lin:exact} built
of~$q$-oscillators~\cite{kli-sch:quantum,maj:foundations}.
Its~$n$-particle Bethe Ansatz eigenfunctions amount to the celebrated Hall--Littlewood
functions~\cite{die-ems:diagonalization,kor:cylindric,tsi:quantum}.
The model in question can moreover be viewed as a~degeneration of the recently found stochastic \mbox{$q$-Hahn} particle
system~\cite{bor-cor-pet-sas:spectral,pov:integrability,tak:discrete}.

{\sloppy By deforming the~$q$-oscillator algebra at the boundary, a~semi-inf\/inite~$q$-boson system was
constructed~\cite{die-ems:boundary, die-ems:semi-infinite} with eigenfunctions given by the hyperoctahedral
Hall--Littlewood func\-tions~\mbox{\cite{mac:orthogonal,ven:symmetric}} that arise as a~($q\to 0$) limit of the
Macdonald--Koorwinder polynomials~\cite{koo:askey-wilson,mac:orthogonal}.
In the present note, we use a~corresponding $q\to 0$ degeneration of the Pieri formulas for the Macdonald--Koornwinder
polynomials~\cite{die:properties} to arrive at explicit formulas for the commuting quantum integrals of the latter
semi-inf\/inite~$q$-boson system with boundary interactions.

}

For the~$q$-boson systems on the f\/inite periodic lattice and on the (bi-)inf\/inite lattice, analogous descriptions of the
commuting quantum integrals stemming from the Pieri formulas for the Hall--Littlewood functions can be found
in~\cite{die:diagonalization,kor:cylindric,tsi:quantum} and in~\cite{die-ems:diagonalization}, respectively.
Previously, Pieri formulas for Macdonald's ($q$-deformed Hall--Littlewood) polynomials were interpreted in a~similar
vein as eigenvalue equations for the quantum integrals of lattice Ruijsenaars--Schneider type
models~\cite{die:scattering,die-vin:quantum,rui:finite-dimensional,rui:factorized}.

\section{Commuting quantum integrals}\label{sec2}

Let $e_1,\ldots,e_n$ be the standard unit basis of $\mathbb{R}^n$ and let~$\Lambda$ denote the cone of integer
partitions $\lambda=(\lambda_1,\ldots,\lambda_n)$ with parts $\lambda_1\geq\dots\geq\lambda_n\geq 0$.

For $l\in \{1,\ldots,n\}$, we def\/ine the following dif\/ference operator $H_l$ acting on the space $C(\Lambda)$ of complex
lattice functions $f\colon \Lambda \to\mathbb{C}$:
\begin{gather}
\label{Hint}
(H_l f)(\lambda):=
\sum\limits_{\substack{J_+, J_-\subset \{1,\ldots,n\}\\J_+\cap J_-=\varnothing, \, |J_+|+| J_- | \leq l\\ \lambda + e_{J_+}-e_{J_-}\in\Lambda}}
U_{J_+^c\cap J_-^c, l-|J_+|-|J_-|}(\lambda) W_{J_+,J_-}(\lambda)
f(\lambda+e_{J_+}-e_{J_-}),
\end{gather}
where $|J|$ refers to the cardinality of $J\subset \{1,\ldots,n\}$, $J^c:=\{1,\ldots,n\}\setminus J$, and $e_J:=\sum\limits_{j\in J} e_j$.
The coef\/f\/icients of the dif\/ference operator in question are built of the factors
\begin{gather*}
W_{J_+,J_-}(\lambda): = \prod\limits_{\substack{j\in J_+\\ \lambda_j=0}}
\left(\prod\limits_{0\leq r<s\leq 2} \big(1-t_rt_st^{n-j}\big) \right) \prod\limits_{\substack{1\leq j < k\leq n\\ \lambda_j=\lambda_k, \, \epsilon_j>\epsilon_k}}
\frac{1-t^{1+k-j}}{1-t^{k-j}},
\end{gather*}
with $\epsilon_j \equiv \epsilon_j(J_+,J_-)$, and
\begin{gather*}
U_{K, m}(\lambda):=(-1)^m
 \sum\limits_{\substack{I_+,I_-\subset K\\I_+\cap I_-=\varnothing, \, |I_+|+|I_-|= m}}
\left(\prod\limits_{\substack{j\in I_+\\ \lambda_j=0}}
\big(1-t_0{t}_1 t^{n-j}\big)\big(1-t_0t_2t^{n-j}\big)\right.
\\
\phantom{U_{K, m}(\lambda):=}{}
\times \prod\limits_{\substack{j\in I_- \\ \lambda_j=1}} \big(1- t_1t_2t^{n-j}\big)
\prod\limits_{\substack{j, k\in K\\ \lambda_j=\lambda_k, \, \epsilon_j>\epsilon_k}}
\frac{1-t^{1+k-j}}{1-t^{k-j}} \prod\limits_{\substack{j\in I_-, \, k\in I_+\\ \lambda_j=\lambda_k+1}} \frac{1-t^{1+k-j}}{1-t^{k-j}}
\\
\left.\phantom{U_{K, m}(\lambda):=}{}
\times \prod\limits_{j\in K} t_0^{-\epsilon_j} \prod\limits_{\substack{j,k\in K, \, j <k\\ \epsilon_j\neq\epsilon_k=0}} t^{-\epsilon_j}
 \prod\limits_{\substack{j,k\in K, \, j <k\\ \lambda_j=\lambda_k, \epsilon_k-\epsilon_j=1}} t^{-1} \right),
\end{gather*}
with $\epsilon_j\equiv \epsilon_j(I_+,I_-)$.
Here we have employed the notation
\begin{gather*}
\epsilon_j(J_+,J_-):=
\begin{cases}
1&\text{if}\quad  j\in J_+,
\\
-1&\text{if}\quad  j\in J_-,
\\
0&\text{otherwise}
\end{cases}
\end{gather*}
for $J_+,J_-\subset\{1,\ldots,n\}$ with $J_+\cap J_-=\varnothing$, and we have also assumed the standard convention that
empty products are equal to~$1$.

When $l=1$, the action of $H_l$~\eqref{Hint} is relatively straightforward.
Indeed, in this special case our operator amounts to a~second-order dif\/ference operator of the form
\begin{gather}
\label{H}
(H_1 f)(\lambda)=u(\lambda)f(\lambda)+\sum\limits_{\substack{1\leq j\leq n\\ \lambda+e_j\in \Lambda}} w^+_j(\lambda)f(\lambda+e_j)
+ \sum\limits_{\substack{1\leq j\leq n\\ \lambda-e_j\in \Lambda}} w^-_j(\lambda)f(\lambda-e_j),
\end{gather}
with
\begin{gather*}
w^+_j(\lambda) = \left(\prod\limits_{0\leq r<s\leq 2} (1-t_rt_st^{n-j}) \right)^{\delta_{\lambda_j}}
\prod\limits_{\substack{j< k \leq n\\ \lambda_k=\lambda_j}} \frac{1-t^{1+k-j}}{1-t^{k-j}},
\qquad
w^-_j(\lambda) = \prod\limits_{\substack{1\leq k <j\\ \lambda_k=\lambda_j}} \frac{1-t^{1+j-k}}{1-t^{j-k}}
\end{gather*}
and
\begin{gather*}
u(\lambda)= - \sum\limits_{\substack{1\leq j\leq n\\ \lambda+e_j\in \Lambda}}
t_0^{-1}t^{-(n-j)}\big(\big(1-t_0{t}_1 t^{n-j}\big)\big(1-t_0t_2t^{n-j}\big)\big)^{\delta_{\lambda_j}}
\prod\limits_{\substack{j< k \leq n\\ \lambda_k=\lambda_j}} \frac{1-t^{1+k-j}}{1-t^{k-j}}
\\
\phantom{u(\lambda)=}{}
 -\sum\limits_{\substack{1\leq j\leq n\\ \lambda-e_j\in \Lambda}} t_0t^{n-j}\big(1- t_1t_2t^{n-j}\big)^{\delta_{\lambda_j-1}}
 \prod\limits_{\substack{1\leq k <j\\ \lambda_k=\lambda_j}} \frac{1-t^{1+j-k}}{1-t^{j-k}},
\end{gather*}
where $\delta_m:=1$ if $m=0$ and $\delta_m:=0$ if $m\neq 0$.
Upon adding a~harmless constant term of the form $\sum\limits_{1\leq j\leq n} (t_0t^{n-j}+t_0^{-1}t^{-(n-j)})$ to the
potential $u(\lambda)$, this reproduces the action of the semi-inf\/inite~$q$-boson Hamiltonian
from~\cite[Section~4.1]{die-ems:boundary} restricted to the~$n$-particle subspace (cf.\ also the proof of~\cite[Proposition~3]{die-ems:boundary}).
The~$q$-boson Hamiltonian considered in~\cite{die-ems:semi-infinite} amounts in turn to the parameter degeneration
$t_2\to 0$ (cf.\ \cite[Section~4.2]{die-ems:boundary}).
In this interpretation the parts of~$\lambda$ represent the positions of~$n$ interacting quantum
particles~-- dubbed~$q$-bosons~-- that hop on a~one-dimensional half-lattice formed by the nonnegative integers.
The parameter~$t$ corresponds to the deformation parameter~$q$ of the underlying~$q$-oscillator algebra and the
parameters~$t_0$,~$t_1$,~$t_2$ play the role of coupling constants for the boundary interaction on the lattice end-point at
the origin (cf.\ Remark~\ref{boundary:rem} at the end)~\cite{die-ems:boundary, die-ems:semi-infinite}.
Here we will generally think of these parameters as rational indeterminates, unless explicitly understood otherwise.

The main result of this note is the following theorem, which states that the dif\/ference operators $H_1,\ldots,H_n$
constitute a~system of commuting quantum integrals for the~$n$-particle~$q$-boson Hamiltonian $H=H_1$~\eqref{H}.
\begin{theorem}
\label{integrability:thm}
The difference operators $H_1,\ldots,H_n$~\eqref{Hint} mutually commute.
\end{theorem}

\section{Proof}

Our proof of Theorem~\ref{integrability:thm} hinges on the $q\to 0$ degeneration of an explicit Pieri formula for the
Macdonald--Koorwinder polynomials from~\cite[Section~6]{die:properties}.

\subsection[Macdonald--Koornwinder polynomials at $q=0$]{Macdonald--Koornwinder polynomials at $\boldsymbol{q=0}$}

In the limit $q\to 0$, the Macdonald--Koornwinder polynomial~\cite{koo:askey-wilson,mac:affine} gives rise to a~two-parameter
extension of Macdonald's $BC$-type Hall--Littlewood function~\cite[\S~10]{mac:orthogonal} of the form~\cite{ven:symmetric}
\begin{gather}
\label{HL}
P_\lambda (\xi) =c_\lambda \sum\limits_{w\in W} C_\lambda(w \xi) e^{-i \langle \lambda, w \xi\rangle},
\qquad
\lambda\in\Lambda,
\end{gather}
with
\begin{gather*}
C_\lambda(\xi):=\prod\limits_{1\leq j < k\leq n} \frac{\big(1-t e^{i(\xi_j-\xi_k)}\big)\big(1-t e^{i(\xi_j+\xi_k)}\big)}{\big(1-e^{i(\xi_j-\xi_k)}\big)\big(1- e^{i(\xi_j+\xi_k)}\big)}
\prod\limits_{\substack{1\leq j \leq n\\ \lambda_j>0}}\frac{\prod\limits_{r=0}^3\big(1-t_r e^{i\xi_j}\big)}{1- e^{2i\xi_j}}
\end{gather*}
and
\begin{gather*}
c_\lambda:= \prod\limits_{1\leq j\leq n}
\frac{t_0^{\lambda_j}t^{(n-j)\lambda_j}(1-t)}{(1-t^j)(1+t^{n-j})^{\delta_{\lambda_j}} \prod\limits_{1\leq r\leq
3}(1-t_0t_rt^{n-j})^{1-\delta_{\lambda_j}}}.
\end{gather*}
Here $\langle\lambda,\xi\rangle \equiv\sum\limits_{j=1}^n \lambda_j\xi_j$ and $W=S_n\ltimes \{-1,1\}^n$ denotes the
hyperoctahedral group of signed permutations acting linearly on the components of
$\xi=(\xi_1,\ldots,\xi_n)\in\mathbb{C}^n$.
The trigonometric polynomial $P_\lambda(\xi)$~\eqref{HL} is normalized so as to achieve a~unit principal specialization
value~\cite{die-ems:boundary}:
\begin{gather*}
P_\lambda (i\rho) =1
\qquad
\text{at}
\quad
\rho =\sum\limits_{1\leq j\leq n} (\log (t_0)+(n-j)\log(t))e_j.
\end{gather*}

For parameter values in the domain
\begin{gather*}
t, t_r \in (-1,1)\setminus \{0 \},
\qquad
r=0,\ldots,3,
\end{gather*}
the $q\to 0$ degeneration of the Macdonald--Koornwinder orthogonality ensures that~\cite{koo:askey-wilson,ven:symmetric}
\begin{gather}
\label{ortho}
\int_{[0,2\pi]^n} P_\lambda (\xi) \overline{P_\mu (\xi)} |\Delta (\xi)|^2\text{d}\xi= 0  
\qquad
\text{if}\quad  \lambda\neq\mu,
\end{gather}
where the orthogonality density is given by the squared modulus of
\begin{gather*}
\Delta (\xi) =\prod\limits_{1\leq j < k\leq n} \frac{\big(1- e^{i(\xi_j-\xi_k)}\big)\big(1- e^{i(\xi_j+\xi_k)}\big)}{\big(1-
te^{i(\xi_j-\xi_k)}\big)\big(1- te^{i(\xi_j+\xi_k)}\big)} \prod\limits_{1\leq j \leq n}\frac{1- e^{2i\xi_j}}
{\prod\limits_{r=0}^3\big(1-t_r e^{i\xi_j}\big)}.
\end{gather*}

\subsection{Pieri formulas}

For a~given choice of generators $E_1(\xi),\ldots,E_n(\xi)$ for the algebra of trigonometric polynomials with
hyperoctahedral symmetry, the associated expansions of products of the form $E_l(\xi)P_\lambda (\xi)$ in the basis
$P_\mu(\xi)$, $\mu\in\Lambda$ give rise to a~system of recurrence relations commonly referred to as Pieri formulas.

\begin{theorem}
\label{pieri:thm}
For $l\in\{1,\ldots,n\}$, the hyperoctahedral Hall--Littlewood function $P_\lambda (\xi)$~\eqref{HL} with $t_3=0$
satisfies the following Pieri-type recurrence relation:
\begin{gather*}
E_l(\xi) P_\lambda (\xi)
=\sum\limits_{\substack{J_+, J_-\subset \{1,\ldots,n\} \\ J_+\cap J_-=\varnothing, \, |J_+|+| J_- | \leq l \\ \lambda + e_{J_+}-e_{J_-}\in\Lambda}}
U_{J_+^c\cap J_-^c, l-|J_+|-|J_-|}(\lambda) V_{J_+,J_-}(\lambda)P_{\lambda+e_{J_+}-e_{J_-}} (\xi),
\end{gather*}
with
\begin{gather*}
E_l (\xi):=\sum\limits_{1\leq j_1<\dots < j_l\leq n} \prod\limits_{1\leq k\leq l}\big(2\cos(\xi_{j_k})-t^{j_k-k}{t}_0-t^{-(j_k-k)}{t}_0^{-1}\big),
\\
V_{J_+,J_-}(\lambda): =  \prod\limits_{\substack{j\in J_+\\ \lambda_j=0}} \big(1-t_0t_1t^{n-j}\big) \big(1-t_0t_2t^{n-j}\big)
\prod\limits_{\substack{j\in J_-\\ \lambda_j=1}} \big(1-t_1t_2t^{n-j}\big)
\\
\hphantom{V_{J_+,J_-}(\lambda): =}{}
\times \prod\limits_{\substack{1\leq j < k\leq n\\ \lambda_j=\lambda_k, \epsilon_j>\epsilon_k}}
\frac{1-t^{1+k-j}}{1-t^{k-j}} \prod\limits_{1\leq j\leq n}
t_0^{-\epsilon_j}t^{-(n-j)\epsilon_j},
\end{gather*}
where $\epsilon_j\equiv \epsilon_j(J_+,J_-)$, and with $U_{K,m}(\lambda) $ being defined as in Section~{\rm \ref{sec2}}.
\end{theorem}

\begin{proof}
The stated formula boils down to a~degeneration of the Pieri formula for the Macdonald--Koornwinder polynomials
in~\cite[Theorem~6.1]{die:properties}.
This degenerate formula is obtained through a~straightforward but somewhat tedious computation that involves setting
$t_3=q$ and performing the limit $q\to 0$.
The present formulation moreover employs a~compact expression for the multiplying polynomial $E_l(\xi)$ stemming
from~\cite[equation~(5.1)]{kom-nou-shi:kernel} (cf.\ also~\cite[Section~2]{die-ems:branching}).
\end{proof}

When $t_2=0$, Theorem~\ref{pieri:thm} reduces to a~Pieri formula for Macdonald's $BC$-type Hall--Littlewood functions.

\subsection{Commutativity}
\label{commutativity:subs}
For $\lambda\in\Lambda$ such that $\lambda+e_{J_+}-e_{J_-}\in\Lambda$, we have the following functional relation between
the coef\/f\/icients of the dif\/ference operators in Theorem~\ref{integrability:thm} and those of the Pieri formulas in
Theorem~\ref{pieri:thm}:
\begin{gather*}
V_{J_+,J_-}(\lambda)h(\lambda+e_{J_+}-e_{J_-})= W_{J_+,J_-}(\lambda)h(\lambda),
\end{gather*}
with
\begin{gather*}
h(\lambda):=\prod\limits_{\substack{1\leq j\leq n\\ \lambda_j >0}} t_0^{\lambda_j}t^{(n-j)\lambda_j}\big(1-t_1t_2t^{n-j}\big).
\end{gather*}
The upshot is that upon conjugating $H_l$~\eqref{Hint} with the (invertible) multiplication operator in~$C(\Lambda)$ of
the form $f\to hf$, the coef\/f\/icients $W_{J_+,J_-}(\lambda)$ get replaced by the Pieri coef\/f\/i\-cients~$V_{J_+,J_-}(\lambda)$.
The Pieri formula in Theorem~\ref{pieri:thm} tells us that the resulting conjugated dif\/ference operators commute on the
joint eigenbasis of hyperoctahedral Hall--Littlewood functions~\eqref{HL} (viewed as lattice functions of~$\lambda\in
\Lambda$ depending on a~polynomial spectral parameter $\xi\in\mathbb{R}^n$).
Indeed, from this perspective the Pieri formula corresponds to an eigenvalue equation with the bounded function
$E_l(\xi)$ playing the role of the eigenvalue.
The orthogonality relations~\eqref{ortho} moreover guarantee the completeness of these (generalized) eigenfunctions
(cf.\ Remark~\ref{completeness:rem} below), i.e., the dif\/ference operators in question commute in fact as bounded
operators in the Hilbert space $\ell^2(\Lambda,\nu)\subset C(\Lambda)$ determined by a~discrete measure with weights
\begin{gather*}
\nu_\lambda = \left(\int_{[0,2\pi]^n} | P_\lambda (\xi) \Delta (\xi)|^2\text{d}\xi \right)^{-1},
\qquad
\lambda\in\Lambda.
\end{gather*}
This means that the operators commute in particular on the (stable) subspace of $C(\Lambda)$ consisting of the lattice
functions with f\/inite support.
But then the commutativity must actually hold on the whole space $C(\Lambda)$, as given any $f\in C(\Lambda)$ and any
$\lambda\in \Lambda$, the evaluation at~$\lambda$ of the commutator of two of such dif\/ference operators acting on~$f$
depends manifestly only on evaluations of~$f$ at a~f\/inite number of lattice points in~$\Lambda$.
Finally, the commutativity is extended beyond the parameter values in the orthogonality domain by analyticity.

\begin{remark}
\label{completeness:rem}
Notice that away from the boundary (i.e., for $\lambda_1\geq\dots\geq\lambda_n>0$) the wave function $P_\lambda
(\xi)$~\eqref{HL} decomposes as a~linear combination of plane waves (of Bethe Ansatz form).
In particular, the wave function in question does {\em not} belong to the Hilbert space $\ell^2(\Lambda,\nu)$ and
constitutes in fact a~{\em generalized} eigenfunction of the discrete dif\/ference operators arising from the Pieri formulas.
(The spectra of these bounded dif\/ference operators are absolutely continuous rather than discrete.) The orthogonality
relations~\eqref{ortho} do nevertheless imply that any~$f$ in $\ell^2(\Lambda,\nu)$ can be represented through a~wave
packet via the associated (generalized) Fourier transform:
\begin{gather*}
f(\lambda)=\int_{[0,2\pi]^n} \hat{f}(\xi) {P_\lambda (\xi)} |\Delta (\xi)|^2\text{d}\xi,
\qquad
\lambda\in \Lambda,
\end{gather*}
where
\begin{gather}
\label{fti}
\hat{f}(\xi) = \sum\limits_{\lambda\in \Lambda} f(\lambda) \overline{P_\lambda (\xi)} \nu_\lambda,
\end{gather}
which conf\/irms the completeness of our generalized eigenfunctions.
(Here the convergence of the sum on the r.h.s.\ of~\eqref{fti} is in the strong $L^2([0,2\pi]^n,|\Delta (\xi)|^2
\text{d}\xi)$ Hilbert space sense.)
\end{remark}

\section{Extension to four-parameter boundary interactions}

In principle there is no genuine obstruction preventing us from adapting the commuting quantum integrals of
Theorem~\ref{integrability:thm} to the case of the more general semi-inf\/inite~$q$-boson system
in~\cite{die-ems:boundary} with four-parameter boundary interactions.
For this purpose, one needs to establish a~ge\-ne\-ra\-lization of the Pieri formula in Theorem~\ref{pieri:thm} covering the
hyperoctahedral Hall--Littlewood polynomials $P_\lambda (\xi)$~\eqref{HL} with $t_3$ arbitrary.
It turns out, however, that in this more general setting the coef\/f\/icients of the Pieri formulas (and thus also those of
the corresponding quantum integrals) become quite baroque.
We wrap up by indicating brief\/ly how the above formulas are to be modif\/ied when dealing with such general boundary
interactions involving four parameters $t_0,\ldots,t_3$.

\subsection{Pieri coef\/f\/icients}
Even though the global structure of the Pieri formula for the hyperoctahedral Hall--Littlewood functions $P_\lambda
(\xi)$~\eqref{HL} remains of the form described by Theorem~\ref{pieri:thm} when dropping the condition that $t_3$ be
zero, the f\/ine structure of the coef\/f\/icients is now more intricate:
\begin{gather*}
V_{J_+,J_-}(\lambda)
=\prod\limits_{\substack{j\in J_+\\ \lambda_j=0}}
\frac{\big(1-\tau t^{n-j+m_0(\lambda)+m_1(\lambda)-m_1^+(\lambda)}\big) \prod\limits_{1\leq r\leq
3}(1-t_0t_rt^{n-j})} {(1-\tau t^{2(n-j)}) (1-\tau t^{2(n-j)+1})}
\\
\phantom{V_{J_+,J_-}(\lambda)=}{}
\times \prod\limits_{\substack{j\in J_+\\ \lambda_j=1}} \big(1-\tau t^{n-j+m_0(\lambda)}\big)
\prod\limits_{\substack{j\in J_-\\ \lambda_j=1}}
\frac{(1-\tau t^{n-j-1})\prod\limits_{1\leq r < s\leq 3}(1-t_rt_st^{n-j})}{(1-\tau t^{2(n-j)}) (1-\tau t^{2(n-j)-1})}
\\
\phantom{V_{J_+,J_-}(\lambda)=}{}
\times \prod\limits_{\substack{1\leq j < k\leq n\\ \lambda_j=\lambda_k, \, \epsilon_j>\epsilon_k}}
\frac{1-t^{1+k-j}}{1-t^{k-j}} \prod\limits_{1\leq j\leq n}t_0^{-\epsilon_j}t^{-(n-j)\epsilon_j}
\end{gather*}
and
\begin{gather}
U_{K, m}(\lambda)=(-1)^m \sum\limits_{\substack{I_+,I_-\subset K\\I_+\cap I_-=\varnothing, \, |I_+|+|I_-|= m}}
\left(\prod\limits_{\substack{j\in I_+\\ \lambda_j=0}}
\frac{\prod\limits_{1\leq r\leq 3}(1-t_0{t}_r t^{n-j})}{1-\tau t^{2(n-j)}}
\prod\limits_{\substack{j\in I_+\\ \lambda_j=1}} (1-\tau t^{n-j})\right.
\nonumber
\\
\phantom{U_{K, m}(\lambda)=}{}
\times\prod\limits_{\substack{j\in I_-\\ \lambda_j=1}}\frac{\prod\limits_{1\leq r <s\leq 3} (1- t_rt_st^{n-j})}{1-\tau t^{2(n-j)}}
\prod\limits_{\substack{j,k\in K, \, j<k\\ \epsilon_j+\epsilon_k\in \{-2,1,2\}\\ \lambda_j=1, \, \lambda_k=\delta_{1+\epsilon_k}}}
\frac{1-\tau t^{2n+1-j-k}}{1-\tau t^{2n-j-k}}
\nonumber
\\
\phantom{U_{K, m}(\lambda)=}{}
\times
\prod\limits_{\substack{j\in I_+\cup I_-,\, k\in K\setminus I_-\\j<k, \, \epsilon_k-\epsilon_j\in\{0,1\}\\ \lambda_j=\delta_{1+\epsilon_j}, \, \lambda_k=0}}
\frac{1-\tau t^{2n-1-j-k}}{1-\tau t^{2n-j-k}}
\prod\limits_{\substack{j, k\in K\\ \lambda_j=\lambda_k,\, \epsilon_j>\epsilon_k}}
\frac{1-t^{1+k-j}}{1-t^{k-j}} \prod\limits_{\substack{j\in I_-, \, k\in I_+\\ \lambda_j=\lambda_k+1}} \frac{1-t^{1+k-j}}{1-t^{k-j}}
\nonumber
\\
\left.\phantom{U_{K, m}(\lambda)=}{}
 \times \prod\limits_{j\in K} t_0^{-\epsilon_j} \prod\limits_{\substack{j,k\in K,\,  j <k\\ \epsilon_j\neq\epsilon_k=0}} t^{-\epsilon_j}
 \prod\limits_{\substack{j,k\in K, \, j <k\\ \lambda_j=\lambda_k, \, \epsilon_k-\epsilon_j=1}} t^{-1} \right),
\label{Ukm}
\end{gather}
where $m_l (\lambda)=|\{1\leq j\leq n \mid \lambda_j=l\} |$, $m_l^+ (\lambda)=|\{j\in J_+ \mid \lambda_j=l\}|$, and $\tau= t_0t_1t_2t_3$.
These Pieri coef\/f\/icients are obtained from~\cite[Theorem 6.1]{die:properties} in the limit $q\to 0$.

\subsection[$q$-Boson quantum integrals]{$\boldsymbol{q}$-Boson quantum integrals}

By virtue of the argumentation in
Section~\ref{commutativity:subs}, the Pieri formulas at issue give rise to commuting dif\/ference operators $H_1,\ldots,
H_n$ of the form stated in equation~\eqref{Hint} with
\begin{gather*}
 W_{J_+,J_-}(\lambda) =\prod\limits_{\substack{j\in J_+\\ \lambda_j=1}} \big(1-\tau t^{n-j+m_0(\lambda)}\big)
\prod\limits_{\substack{1\leq j < k\leq n\\ \lambda_j=\lambda_k, \, \epsilon_j>\epsilon_k}} \frac{1-t^{1+k-j}}{1-t^{k-j}}
\\
\phantom{ W_{J_+,J_-}(\lambda)=}{}
\times \prod\limits_{\substack{j\in J_+\\ \lambda_j=0}}
\frac{\big(1-\tau t^{n-j-1}\big)\big(1-\tau t^{n-j+m_0(\lambda)+m_1(\lambda)-m_1^+(\lambda)}\big) \prod\limits_{0\leq
r<s\leq 3} \big(1-t_rt_st^{n-j}\big)}{(1-\tau t^{2(n-j)-1}) (1-\tau t^{2(n-j)})^2 (1-\tau t^{2(n-j)+1})}
\end{gather*}
and with $U_{K,m}(\lambda)$ taken from equation~\eqref{Ukm}.
The relation between the coef\/f\/icients of the Pieri formula and those of the dif\/ference operators is again governed~by
a~functional identity of the type in Section~\ref{commutativity:subs} with
\begin{gather*}
h(\lambda)=\prod\limits_{1\leq j\leq n} t_0^{\lambda_j}t^{(n-j)\lambda_j} \big(1-\tau
t^{n+m_0(\lambda)-j-1}\big)^{\delta_{\lambda_j}} \prod\limits_{1\leq r<s\leq 3} \big(1-t_rt_st^{n-j}\big)^{1-\delta_{\lambda_j}}.
\end{gather*}
When $l=1$ the corresponding dif\/ference operator $H_l$ now reduces to $H_1$~\eqref{H} with
\begin{subequations}
\begin{gather}
w^+_j(\lambda) = \big(1-\tau t^{2m_0(\lambda)+m_1(\lambda)-1}\big)^{\delta_{\lambda_j-1}+\delta_{\lambda_j}}
\prod\limits_{\substack{j< k \leq n \\ \lambda_k=\lambda_j}} \frac{1-t^{1+k-j}}{1-t^{k-j}}
\nonumber
\\
\phantom{w^+_j(\lambda) =}
 \times \left(\frac{(1-\tau t^{n-j-1}) \prod\limits_{0\leq r<s\leq 3} (1-t_rt_st^{n-j})}{(1-\tau t^{2(n-j)-1})(1-\tau
t^{2(n-j)})^2 (1-\tau t^{2(n-j)+1})} \right)^{\delta_{\lambda_j}},
\label{w+}
\\
w^-_j(\lambda) = \prod\limits_{\substack{1\leq k <j\\ \lambda_k=\lambda_j}} \frac{1-t^{1+j-k}}{1-t^{j-k}},
\end{gather}
and
\begin{gather}
u(\lambda)=- \sum\limits_{\substack{1\leq j\leq n\\ \lambda+e_j\in \Lambda}}
t_0^{-1}t^{-(n-j)}\big(1-\tau t^{2m_0(\lambda)+m_1(\lambda)-1}\big)^{\delta_{\lambda_j-1}+\delta_{\lambda_j}}
\nonumber
\\
\phantom{u(\lambda)=- \sum\limits_{\substack{1\leq j\leq n\\ \lambda+e_j\in \Lambda}}}{}
 \times \left(\frac{\prod\limits_{1\leq r\leq 3} (1-t_0{t}_r t^{n-j})}{(1-\tau t^{2(n-j)})(1-\tau t^{2(n-j)+1})}
\right)^{\delta_{\lambda_j}} \prod\limits_{\substack{j< k \leq n\\ \lambda_k=\lambda_j}} \frac{1-t^{1+k-j}}{1-t^{k-j}}
\label{u}
\\
\phantom{u(\lambda)=}{}
-\sum\limits_{\substack{1\leq j\leq n\\ \lambda-e_j\in \Lambda}}
t_0t^{n-j}\left(\frac{(1-\tau t^{n-j-1})\prod\limits_{1\leq r<s\leq 3}(1- t_rt_st^{n-j})}
{(1-\tau t^{2(n-j)-1})(1-\tau t^{2(n-j)})} \right)^{\delta_{\lambda_j-1}}
\prod\limits_{\substack{1\leq k <j\\ \lambda_k=\lambda_j}} \frac{1-t^{1+j-k}}{1-t^{j-k}}.
\nonumber
\end{gather}
\end{subequations}
Upon adding the constant term $\sum\limits_{1\leq j\leq n} (t_0t^{n-j}+t_0^{-1}t^{-(n-j)})$, this reproduces
the~$n$-particle~$q$-boson Hamiltonian of~\cite[Section 3]{die-ems:boundary}.

\begin{remark}
\label{boundary:rem}
In~\cite{die-ems:boundary} a~Fock space description of the particle Hamiltonian $H_1$~\eqref{H},~\eqref{w+}--\eqref{u}
was provided as a~system of~$q$-bosons on the nonnegative integer lattice perturbed at the lattice-end.
Specif\/ically, the boundary interactions arise in this picture from a~deformation of the~$q$-boson f\/ield algebra at the
origin and its nearest neighboring lattice point parametrized by $t_0$, $t_1$, $t_2$, $t_3$.
In general, i.e., when $\tau=t_0t_1t_2t_3\neq 0$, the deformed~$q$-boson f\/ield algebra is no longer ultralocal at the
boundary as the commutativity between the creation and annihilation operators at the origin and its nearest neighboring site is lost.
When $t_3=0$, the deformation of the~$q$-boson f\/ield algebra is restricted only to the lattice end-point at the origin
and the ultralocality is restored~\cite[Section~4.1]{die-ems:boundary}.
When both $t_2=t_3=0$, the boundary interaction degenerates and decomposes into an interaction arising from
a~one-parameter deformation of the~$q$-boson f\/ield algebra and a~one-parameter additive potential term of the
Hamiltonian, supported at the lattice end-point~\cite{die-ems:semi-infinite}.
\end{remark}

\subsection*{Acknowledgements}

This work was supported in part by the {\em Fondo Nacional de Desarrollo Cient\'{\i}f\/ico y Tecnol\'ogico (FONDECYT)}
Grants \# 1130226 and \# 1141114.

\pdfbookmark[1]{References}{ref}
\LastPageEnding

\end{document}